\documentclass[12pt]{article}
\usepackage{enumerate}
\usepackage{amsmath}
\usepackage{amsfonts}
\usepackage{amssymb}
\newcommand{\bt}{\begin{theorem}}
\newcommand{\wt}{\widetilde}

\newcommand{\et}{\end{theorem}}
\newcommand {\be}{\begin{equation}}
\newcommand {\ee}{\end{equation}}

\newcommand{\hsp}{\hskip 2em}

\newcommand{\noi}{\noindent}

\def \qed {\hfill \vrule height6pt width6pt depth0pt}

\title{$k$th price auctions and Catalan numbers}

\begin{document}
 \author{\textsc{Abdel-Hameed Nawar}{\footnote{Faculty of Economics and Political Science, Cairo University, Giza 12613, Egypt.\vspace{0.2cm}}}\hsp
\textsc{Debapriya Sen}{\footnote{Department of Economics, Ryerson University, Toronto, Ontario, Canada.\vspace{0.2cm}}}}

\date{\today}

\maketitle

\centerline{\bf Abstract} \noi This paper establishes an interesting link between $k$th price auctions and Catalan numbers by showing that for distributions that have linear density, the bid function at any symmetric, increasing equilibrium of a $k$th price auction with $k\geq 3$ can be represented as a finite series of $k-2$ terms whose $\ell$th term involves the $\ell$th Catalan number. Using an integral representation of Catalan numbers
together with some classical combinatorial identities, we derive the closed form of the unique symmetric, increasing equilibrium of a $k$th price auction for a non-uniform distribution.

\vspace{0.2cm} \noi{\bf Keywords:} $k$th price auction; the revenue equivalence principle; Catalan numbers; Jensen's identity; Hagen-Rothe's identity

\newpage

\begin{quote}
\small{``Catalan numbers: an integer sequence that materializes in unexpected places"} \\
|Martin Gardner (1976)
\end{quote}

\section{Introduction}

In a $k$th price auction with $k$ or more bidders, the highest bidder wins the object and pays the $k$th highest bid as price. This paper establishes an interesting link of such auctions with Catalan numbers\footnote{Catalan numbers are named after Belgian mathematician Eug\`{e}ne Charles Catalan (1814-1894).
Early works in relation to this sequence can be traced back to Mongolian mathematician Ming'antu (1692-1763). See Pak (2015) for a history and
Stanley (2015) for a comprehensive overview. Also see sequence A000108 of ``The On-line Encyclopedia of Integer Sequences" of Sloane: {\texttt{https://oeis.org/A000108}}\vspace{0.1cm}} by showing that for certain distributions, the bid function at any symmetric, increasing equilibrium of a $k$th price auction with $k\geq 3$ can be represented as a finite series involving Catalan numbers. Using an integral representation of Catalan numbers together with some classical combinatorial identities, we are then able to characterize equilibrium bids and obtain their bounds.

There are results on the existence of equilibrium for $k$th price auctions (e.g., Kagel and Levin, 1993; Monderer and Tennenholtz, 2000), but beyond uniform distributions, closed form expressions of equilibrium bids are mostly unknown. This paper shows that Catalan numbers can help us to fill this void.

We consider an independent private value $k$th price auction with $k\geq 3$, in which there are $k$ or more risk neutral bidders, where values are continuously distributed in a finite interval. To identify symmetric, increasing equilibrium of this auction, we follow the approach of appealing to the revenue equivalence principle.\footnote{Vickrey (1961) introduced independent private value auctions and subsequently (Vickrey, 1962) established revenue equivalence between first and second price auctions. Riley and Samuelson (1981) and Myerson (1981) established the revenue equivalence principle. For a comprehensive presentation of auction theory, see the book of Krishna (2002). There is a small literature on $k$th price auction with complete information (e.g., Tauman, 2002; Mathews and Schwartz, 2017).} The revenue equivalence principle implies
that at any symmetric, increasing equilibrium of a $k$th price auction, for any value the expected payment of a bidder is the same as its expected payment in a second-price auction. Given this
result, the first step would be to see if we can find an expression of bid function using the relation on expected payments. If that can be found, the second step would be to verify if
the resulting bid function is increasing. Provided that is the case, we can conclude that this bid function constitutes a symmetric equilibrium of the $k$th price auction.

For a $k$th price auction with general distributions, the problem of determining the bid function from the revenue equivalence principle is quite involved. We are able to find a closed
form expression of the bid function for distributions that have a linear density function, so that its second and higher order derivatives are all
zero. For these distributions, we show that the bid function is a finite series of $k-2$ terms whose $\ell$th term involves the $\ell$th Catalan number (Lemma 2, Section 3.1). Building on this result, we use the integral representation of Catalan numbers derived by Penson and Sixdeniers (2001), together with the combinatorial identities of Jensen and Hagen-Rothe to show that for a specific non-uniform distribution (triangle distribution), the resulting bid function is increasing. This shows that the strategy profile where all bidders follow this bid function is the unique symmetric, increasing equilibrium of the $k$th price auction (Theorem 1, Section 3.3). We also obtain simple lower and upper bounds of the equilibrium bid function.

The paper is organized as follows. We present the basic framework of a $k$th price auction in Section 2. The analysis of equilibrium bids is presented in Section 3. Some proofs are presented in the Appendix.

\section{The basic framework}

The basic framework draws on Chapters 2 and 3 of Krishna (2002). Results that will be useful for our analysis
are summarized here to keep our presentation self contained.

A single object is for sale in an auction. The set of bidders who bid for the object is $N=\{1,\ldots,n\}.$ For $i\in N,$ let $X_i$ be the value of the object for bidder $i.$ Each $X_i$ is independently and identically distributed (iid) on the interval $[0,\omega]$ where $0<\omega<\infty,$ each having an increasing distribution function $F$ that has a continuous density $f\equiv F^\prime$ and has full support. Bidder $i$ knows the realization $x_i$ of $X_i,$ but it only knows that other bidders' values are iid, each following distribution $F.$ Bidders
simultaneously place bids.  An auction is {\it standard} if the rules of the auction are such that the highest bidder wins the object and the payment the winner has to make depends solely
on the submitted bids.

Let $n\geq k\geq 2.$ In a {$k$th price auction}, the highest bidder wins the object and pays the $k$th highest bid. Thus, a $k$th price auction is a standard auction.
A key result that will be useful for our analysis is the revenue equivalence principle, which  holds for any standard auction (see Proposition 3.1 of Krishna, 2002).

A $k$th price auction results in a game among the $n$ bidders where the strategy for bidder $i$ is a function $\beta^i:[0,\omega]\rightarrow \mathbb{R}_+$ which determines
its bid for any value. A strategy profile specifies the strategy of each bidder, so it is given by $(\beta^1,\ldots,\beta^n).$ A strategy profile is {\it symmetric} if all bidders have the same strategy in that profile. A symmetric strategy profile is {\it increasing} if the common strategy of that profile is an increasing function.

Fix any bidder: say bidder $n.$ For $r=1,\ldots,n-1,$ let $Y_r$ denote the $r$th highest value among the remaining $n-1$ bidders, that is, $Y_r$ is the $r$th highest order statistic of $X_1,\ldots,X_{n-1}.$ In particular, $Y_1=\max\{X_1,\ldots,X_{n-1}\}.$ Denoting by $G$ the distribution function and $g$ the density function of $Y_1,$ we have
\be\label{y1}G(y)=F(y)^{n-1}\mbox{ and }g(y)=(n-1)F(y)^{n-2}f(y)\ee
The revenue equivalence principle implies that a symmetric, increasing strategy profile is an equilibrium of a $k$th price auction if and only if at that profile, for any value the expected payment of any
bidder is the same as its expected payment in a second price auction. This result is formally stated in the next proposition.

\vspace{0.2cm}\noi {\bf Proposition 1} {\it For a $k$th price auction with $n$ risk neutral bidders where $n\geq k\geq 2,$ consider a symmetric, increasing strategy profile at which the common strategy of all bidders is the function $\beta:[0,\omega]\rightarrow \mathbb{R}_+.$ Denote by $m^\beta(x)$ the expected payment of a bidder with value $x$ at this profile. Then} (i) $m^\beta(0)=0$ {\it and} (ii) {\it this strategy profile is an equilibrium of the $k$th price auction if and only if}
\be\label{m}m^\beta(x)=\int_0^xyg(y)\mbox{d}y\ee

\vspace{0.2cm}\noi {\bf Proof} See the Appendix. \qed

\vspace{0.2cm}\noi {\bf Remark 1} Proposition 1 is closely related to Proposition 3.1 of Krishna (2002). The latter is a more general result which applies to all standard auctions (for example, they include all-pay-auctions) and there the condition $m^\beta(0)=0$ is an assumption. Proposition 1 is specific to $k$th price auctions, which enables us to get $m^\beta(0)=0$ as a result.

\section{Equilibrium of a $k$th price auction}

Consider a symmetric, increasing strategy profile for a $k$th price auction at which the common strategy of all bidders is $\beta:[0,\omega]\rightarrow \mathbb{R}_+.$ Consider a specific bidder,
say bidder $n.$ Let $x\in (0,\omega].$ By the monotonicity of $\beta,$ when bidder $n$ has value $x,$ at this strategy profile $\mbox{Pr}(\mbox{bidder }n\mbox{ wins})=\mbox{Pr}(Y_1<x)=G(x).$
As the auction is $k$th price, when bidder $n$ wins it has to pay the $(k-1)$th highest of the remaining bids. So at this profile the expected payment of a bidder who has value $x$ is given by
\be\label{y11}m^\beta(x)=\mbox{Pr}(Y_1<x)E(\beta(Y_{k-1})|Y_1<x)+\mbox{Pr}(Y_1\geq x)0=G(x)E(\beta(Y_{k-1})|Y_1<x)\ee For $y\leq x,$ the density of $Y_{k-1}$ conditional on $Y_1<x$ is given by
\be\label{y12}h_{k-1}(y|Y_1<x)=\frac{n-1}{G(x)}{n-2\choose k-2}[F(x)-F(y)]^{k-2}F(y)^{n-k}f(y)\ee
See Lemma A1 in the Appendix for the derivation of the conditional density of order statistics. Taking $m=n-1$ and $r=k-1$ there gives (\ref{y12}).
By (\ref{y11}) and (\ref{y12}) we have
$$m^\beta(x)=G(x)\int_0^x \beta(y)h_{k-1}(y|Y_1<x)\mbox{d}y$$\be\label{y13}=(n-1){n-2\choose k-2}\int_0^x \beta(y)[F(x)-F(y)]^{k-2}F(y)^{n-k}f(y)\mbox{d}y \ee
Using (\ref{y1}) in (\ref{m}) of Proposition 1 and by (\ref{y13}), we conclude that for a $k$th price auction, a symmetric, increasing strategy profile with
common strategy $\beta_k:[0,\omega]\rightarrow \mathbb{R}_+$ is an equilibrium if and only if for all $x\in [0,\omega],$ the following\footnote{Note that for $x=0,$ both sides of (\ref{y0})
equal zero.} hold:
\be\label{y0}{n-2\choose k-2}\int_0^x \beta_k(y)[F(x)-F(y)]^{k-2}F(y)^{n-k}f(y)\mbox{d}y=\int_0^x yF(y)^{n-2}f(y)\mbox{d}y\ee
Using (\ref{y0}), if we can determine $\beta_k$ and show that it is an increasing function, then we can conclude that it constitutes a symmetric, increasing equilibrium of
the $k$th price auction. To this end, denote$$\phi_0(x):=\int_{0}^{x} \beta_k(y)[F(x)-F(y)]^{k-2}F(y)^{n-k}f(y)\mbox{d}y,$$\be\label{e0}\psi_0(x):=\int_0^xyF(y)^{n-2}f(y)\mbox{d}y\ee Then by (\ref{y0}), we have
${n-2\choose k-2}\phi_0(x)=\psi_0(x).$ Iteratively define
\be\label{e1}\phi_{t+1}(x):=\frac{\phi^\prime_t(x)}{f(x)}\mbox{ and }\psi_{t+1}(x):=\frac{\psi^\prime_t(x)}{f(x)}\mbox{ for }t=0,1,\ldots\ee From (\ref{y0})-(\ref{e1}), it follows that
\be\label{e2}{n-2\choose k-2}\phi_t(x)=\psi_t(x)\mbox{ for }t=0,1,\ldots\ee

\vspace{0.1cm} \noi {\bf Lemma 1} {\it Let $n\geq 3.$ For $k=3,\ldots,n,$ the following hold:}
\be\label{e8}\phi_{k-1}(x)=(k-2)!\beta_{k}(x)F(x)^{n-k}\ee

\vspace{0.1cm} \noi {\bf Proof} See the Appendix. \qed

\vspace{0.1cm} From (\ref{e2}) and (\ref{e8}) it follows that
\be\label{e10}\beta_k(x)=\frac{\psi_{k-1}(x)}{{n-2\choose k-2}(k-2)!F(x)^{n-k}}\ee

\vspace{0.2cm} \noi {\bf Remark 2}  Since $F^\prime(x)=f(x),$ from (\ref{e0}) and (\ref{e1}), we have $\psi_1(x)=xF(x)^{n-2}$ and $\psi_2(x)=x(n-2)F(x)^{n-3}+F(x)^{n-2}/f(x).$ Then by (\ref{e10}) we have
\be\label{k3}\beta_2(x)=x,\hspace{0.1cm}\beta_3(x)=x+\frac{F(x)}{(n-2)f(x)}\ee
As shown in Proposition C of Monderer and Tennenholtz (2000) and Proposition 3.2 of Krishna (2002), if $F$ is log-concave, then $F/f$ is an increasing function and so is $\beta_3.$
In that case the third price auction has a unique symmetric, increasing equilibrium where the common strategy of each bidder is $\beta_3$ given
in (\ref{k3}).

\vspace{0.1cm} In general, for $k\geq 3,$ we are able to determine $\psi_{k-1}$ for distributions where the density function $f$ is linear, so that its first derivative is a constant and all derivatives of second or higher order are zero. For such distributions we obtain a closed form expression of $\psi_{k-1}$ in terms of Catalan numbers.
Using (\ref{e10}), we can then also represent $\beta_k$ in terms of Catalan numbers.

\subsection{Bid function in terms of Catalan numbers}

For non-negative integers $\ell=0,1,\ldots,$ the $\ell$th Catalan number is given by
\be\label{c1}C_\ell=\frac{1}{\ell+1}{2\ell\choose \ell}\ee In particular, $C_0=C_1=1.$ Note that for $\ell=1,2,\ldots,$ Catalan numbers satisfy the recurrence relation
\be\label{c11}C_{\ell}=\frac{2(2\ell-1)}{\ell+1}C_{\ell-1}\ee
Let $n\geq 3.$ For $k=3,\ldots,n$ and $\ell=0,1,\ldots,k-3,$ define
\be\label{tg}\theta^{k}_{\ell}:={n-2\choose k-3-\ell}\frac{C_{\ell}}{2^{\ell}}\ee
From (\ref{c11}), the following recurrence relation holds for $\ell=1,\ldots,k-3$:
\be\label{tg1}\theta^{k+1}_{\ell}=\frac{2\ell-1}{\ell+1}\theta^{k}_{\ell-1}\ee
Since $(n-k+\ell+1){n-2\choose k-3-\ell}=(k-2-\ell){n-2\choose k-2-\ell},$ we have
\be\label{tg2}(n-k+\ell+1)\theta^{k}_{\ell}=(k-2-\ell)\theta^{k+1}_{\ell}\ee The relations (\ref{tg1}) and (\ref{tg2}) will be useful for our analysis. Now we are in a position to state the result that represents $\beta_k(x)$ in terms of Catalan numbers.

\vspace{0.2cm} \noi {\bf Lemma 2} {\it Let $n\geq 3.$ Suppose $f^\prime(x)=a,$ where $a$ is a constant. Then the following hold for $k=3,\ldots,n$}:
\be\label{gen}\frac{\psi_{k-1}(x)}{(k-2)!}={n-2\choose k-2}xF(x)^{n-k}+\sum_{\ell=0}^{k-3}(-1)^{\ell}\theta^k_\ell a^{\ell}\frac{F(x)^{n-k+\ell+1}}{f(x)^{2\ell+1}}\ee
{\it where $\theta^k_\ell$ is given by $(\ref{tg}).$ Consequently, if $\beta_k(x)$ satisfies} (\ref{e10}), {\it then}
\be\label{b0}\beta_k(x)=x+\frac{1}{{n-2 \choose k-2}}\sum_{\ell=0}^{k-3}(-1)^{\ell}\theta^k_\ell a^{\ell}\frac{F(x)^{\ell+1}}{f(x)^{2\ell+1}}\ee

\vspace{0.2cm} \noi {\bf Proof} Observe that (\ref{b0}) will be immediate from (\ref{gen}) by applying (\ref{e10}). We prove (\ref{gen}) by induction on $k.$ First let $k=3.$ Since $\psi_1(x)=xF(x)^{n-2},$ we have
$\psi_2(x)/0!=\psi_2(x)=\psi^\prime_1(x)/f(x)=(n-2)xF(x)^{n-3}+F(x)^{n-2}/f(x),$ which equals the right side of (\ref{gen}) for $k=3$ (since $\theta^3_0=1$). This shows the result holds for
$k=3.$ In what follows, we show that if the result holds for $k,$ it also holds for $k+1.$

Suppose $\psi_{k-1}(x)/(k-2)!$ is given by (\ref{gen}). Denote the two terms on the right side of (\ref{gen}) by $\tau_1(x),\tau_2(x).$ Then $\psi_{k-1}(x)/(k-2)!=\tau_1(x)+\tau_2(x).$
Since $\psi_k(x)=\psi^\prime_{k-1}(x)/f(x),$ we have
\be\label{2}\frac{\psi_k(x)}{(k-1)!}=\frac{\tau^\prime_1(x)+\tau^\prime_2(x)}{(k-1)f(x)}\ee Note from (\ref{gen}) that
$$\frac{\tau^\prime_1(x)}{f(x)}={n-2\choose k-2}x(n-k)F(x)^{n-k-1}+{n-2\choose k-2}\frac{F(x)^{n-k}}{f(x)}$$
\be\label{t2}=(k-1){n-2\choose k-1}xF(x)^{n-k-1}+\theta^{k+1}_0\frac{F(x)^{n-k}}{f(x)}\ee
Now consider the second term of (\ref{gen}). Since $f^\prime(x)=a,$ we have
$$\frac{\tau^\prime_2(x)}{f(x)}=\sum_{\ell=0}^{k-3}(-1)^{\ell}\theta^k_\ell a^{\ell}\frac{(n-k+\ell+1)F(x)^{n-k+\ell}}{f(x)^{2\ell+1}}
$$$$+\sum_{\ell=0}^{k-3}(-1)^{\ell}\theta^k_\ell a^{\ell+1}\frac{(-1)(2\ell+1)F(x)^{n-k+\ell+1}}{f(x)^{2\ell+3}}$$
Note from (\ref{tg2}) that $(n-k+\ell+1)\theta^k_\ell=(k-2-\ell)\theta^{k+1}_\ell.$ Using this and denoting $j=\ell+1$ in the second sum of above, the expression above equals
$$\sum_{\ell=0}^{k-3}(-1)^{\ell}(k-2-\ell)\theta^{k+1}_\ell a^{\ell}\frac{F(x)^{n-k+\ell}}{f(x)^{2\ell+1}}
+\sum_{j=1}^{k-2}(-1)^j(2j-1)\theta^k_{j-1} a^{j}\frac{F(x)^{n-k+j}}{f(x)^{2j+1}}$$
By (\ref{tg1}), $(k-2-\ell)\theta^{k+1}_\ell+(2\ell-1)\theta^k_{\ell-1}=(k-1)\theta^{k+1}_\ell.$ Using this, the expression above equals
$$(k-2)\theta^{k+1}_0\frac{F(x)^{n-k}}{f(x)}+(k-1)\sum_{\ell=1}^{k-3}(-1)^{\ell}\theta^{k+1}_\ell a^{\ell}\frac{F(x)^{n-k+\ell}}{f(x)^{2\ell+1}}
$$$$+(-1)^{k-2}(2k-5)\theta^k_{k-3}a^{k-2}\frac{F(x)^{n-2}}{f(x)^{2k-3}}$$
Taking $\ell=k-2$ in (\ref{tg1}), we have $(2k-5)\theta^k_{k-3}=(k-1)\theta^{k+1}_{k-2}.$ Using this in the expression above, we have
\be\label{3}\frac{\tau^\prime_2(x)}{f(x)}=(k-2)\theta^{k+1}_0\frac{F(x)^{n-k}}{f(x)}
+(k-1)\sum_{\ell=1}^{k-2}(-1)^{\ell}\theta^{k+1}_\ell a^\ell\frac{F(x)^{n-k+\ell}}{f(x)^{2\ell+1}}\ee
From (\ref{2}), (\ref{t2}) and (\ref{3}), we have
$$\frac{\psi_k(x)}{(k-1)!}={n-2\choose k-1}xF(x)^{n-k-1}+\sum_{\ell=0}^{k-2}(-1)^{\ell}\theta^{k+1}_\ell a^{\ell}\frac{F(x)^{n-k+\ell}}{f(x)^{2\ell+1}}$$
This shows if the result holds for $k,$ it also holds for $k+1.$ Since the result holds for $k=3,$ we conclude that the result holds for all $k=3,\ldots,n.$ \qed

\vspace{0.1cm} \noi {\bf Remark 3} On the basis of Lemma 2, we cannot conclude that $\beta_k$ obtained in (\ref{b0}) constitutes an equilibrium of the $k$th price auction. Such a conclusion can be made only when
the resulting $\beta_k$ is an increasing function. Consider the case of a uniform distribution on $[0,\omega].$ Then $F(x)=x/\omega,$ $f(x)=1/\omega$ and $f^\prime(x)=0.$ Taking $a=0$
in (\ref{b0}) and using (\ref{tg}), in that case we have
\be\label{bu}\beta_k(x)=x+\frac{1}{{n-2 \choose k-2}}\frac{\theta^k_0F(x)}{f(x)}=x+\frac{{n-2 \choose k-3}}{{n-2 \choose k-2}}x=x+\frac{k-2}{n-k+1}x\ee
Since $\beta_k$ given in (\ref{bu}) is indeed an increasing function for $n\geq k\geq 2,$ we conclude that when values are iid and uniformly distributed, then the strategy profile with common strategy $\beta_k$ given by
(\ref{bu}) is the unique symmetric increasing equilibrium of the $k$th price auction. This result was obtained in Kagel and Levin (1993, p.878).

\vspace{0.1cm} Since $F(0)=0,$ the general form of distribution function on $[0,\omega]$ that has $f^\prime(x)=a$ is $F(x)=ax^2/2+bx.$ For non-uniform distributions (i.e., $a\neq 0$), we are able to
verify monotonicity of $\beta_k$ given in (\ref{b0}) (and consequently conclude $\beta_k$ constitutes an equilibrium of the $k$th price auction) for distributions where $b=0,$ that is, when
$F(x)=ax^2/2.$ We obtain this result by two key techniques: (i) using an integral representation of Catalan numbers and (ii) applying some classical combinatorial identities.
Let us first state the results on Catalan numbers and combinatorics that will be used in our analysis.

\subsection{Results on Catalan numbers and combinatorics}
Catalan numbers have the following integral representation (see equation (10) of Penson and Sixdeniers, 2001):
\be\label{cint}C_\ell=\frac{1}{2\pi}\int_{0}^{4}u^\ell\sqrt{\frac{4-u}{u}}\mbox{d}u=\frac{2^{2\ell+1}}{\pi}\int_{0}^{1}t^\ell\sqrt{\frac{1-t}{t}}\mbox{d}t\ee where the second equality follows
by substituting $t=u/4.$

Now we state three fundamental combinatorial identities. For our purpose, in all of these identities, $s$ is any non-negative integer, $m$ is any positive real number and $r,z$ are
any real numbers. The first is  {\it Jensen's identity} (see equation (4.1) of Gould and Quaintance, 2010; equation (1) of Guo, 2011):
\be\label{jen}\sum_{\ell=0}^s{m+z\ell\choose \ell}{r-z\ell \choose s-\ell}=\sum_{\ell=0}^s{m+r-\ell \choose s-\ell}z^{\ell}\ee
The second is {\it Hagen-Rothe's identity} (see equation (17) of Gould, 1956; equation (2) of Chu, 2010):\footnote{Gould (1956) proves a more general result. Taking $p=y,$ $q=-\beta$
in equation (17) of Gould (1956) gives the identity of (\ref{hr}).}\be\label{hr}\sum_{\ell=0}^s\frac{m}{m+z\ell}{m+z\ell\choose \ell}{r-z\ell \choose s-\ell}={m+r\choose s}\ee
The third identity is related to Jensen's identity (see the first equation in p.204, Guo, 2011):
\be\label{hr1}\sum_{\ell=0}^s{r-\ell\choose s-\ell}z^\ell=\sum_{\ell=0}^s{r+1\choose s-\ell}(z-1)^\ell\ee

\subsection{Equilibrium for a non-uniform distribution}

Let $F(x)=ax^2/2,$ where $a$ is a positive constant. Then $f(x)=ax$ and $f^\prime(x)=a>0,$ so the condition of Lemma 2 holds. In this case we have
$a^\ell F(x)^{\ell+1}/f(x)^{2\ell+1}=(1/2)^{\ell+1}x$ and by (\ref{b0}) it follows that
\be\label{bnu0}\beta_k(x)=x+\frac{x}{{n-2 \choose k-2}}\sum_{\ell=0}^{k-3}(-1)^{\ell}\frac{\theta^k_\ell}{2^{\ell+1}}\ee
Thus in this case $\beta_k(x)$ is linear with $\beta_k(0)=0.$ We show that the function $\beta_k$ is increasing and therefore it is an equilibrium of the $k$th price auction.
We show that the function $\beta_k$ is increasing and therefore it is an equilibrium of the $k$th price auction.

\vspace{0.2cm} \noi {\bf Theorem 1} {\it Let $n\geq 3$ and $k=3,\ldots,n.$ Consider a $k$th price auction with $n$ risk neutral bidders. Suppose values are iid on $[0,\omega]$ with distribution
function $F(x)=ax^2/2$ where $a>0.$ Then the following hold.}

\begin{enumerate}[(i)]

\item {\it The strategy profile where each bidder has the common strategy $\beta_k:[0,\omega]\rightarrow \mathbb{R}_+$ given by} (\ref{bnu0})
{\it is the unique symmetric, increasing equilibrium of the $k$th price auction.}

\item {\it If $n$ is sufficiently large compared to $k$ $($specifically, $n+4>2k),$ then}
\be\label{bnu1}x+\frac{k-2}{2(n-2)}x\leq \beta_k(x)\leq x+\frac{7(k-2)}{8(n-2)}x\ee

\end{enumerate}

\noi {\bf Proof} (i) Since for $F(x)=ax^2/2,$ the function $\beta_k$ given in (\ref{bnu0}) is the unique solution to (\ref{e10}), if we can show  $\beta_k$ is an increasing
function, it will prove that the unique symmetric, increasing equilibrium of the $k$th price auction has the common strategy $\beta_k$ for each bidder. To show this, denote
\be\label{om}\Omega_k:=\sum_{\ell=0}^{k-3}(-1)^{\ell}\frac{\theta^k_\ell}{2^{\ell+1}}=\sum_{\ell=0}^{k-3}(-1)^{\ell}{n-2\choose k-3-\ell}\frac{C_\ell}{2^{2\ell+1}}\ee
Note from (\ref{bnu0}) that
\be\label{bnu}\beta_k(x)=x+\frac{x}{{n-2 \choose k-2}}\Omega_k\ee
Using the integral representation (\ref{cint}) in (\ref{om}) and then making the transformation $z=1-t,$ we have
$$\Omega_k=\frac{1}{\pi}\sum_{\ell=0}^{k-3}{n-2\choose k-3-\ell}\left[\int_{0}^{1}(-t)^\ell\sqrt{\frac{1-t}{t}}\mbox{d}t\right]$$$$
=\frac{1}{\pi}\sum_{\ell=0}^{k-3}{n-2\choose k-3-\ell}\left[\int_{0}^{1}(z-1)^\ell\sqrt{\frac{z}{1-z}}\mbox{d}z\right]$$
Switching the orders of summation and integration we have
\be\label{om1}\Omega_k=\frac{1}{\pi}\int_0^1\sqrt{\frac{z}{1-z}}\left[\sum_{\ell=0}^{k-3}{n-2\choose k-3-\ell}(z-1)^\ell\right]\mbox{d}z\ee
Taking $r=n-3,$ $s=k-3$ in (\ref{hr1}) we have
\be\label{hr2}\sum_{\ell=0}^{k-3}{n-2\choose k-3-\ell}(z-1)^\ell=\sum_{\ell=0}^{k-3}{n-3-\ell\choose k-3-\ell}z^\ell\ee
By (\ref{om1}) and (\ref{hr2}) we have
\be\label{om2}\Omega_k=\frac{1}{\pi}\int_0^1\sqrt{\frac{z}{1-z}}\left[\sum_{\ell=0}^{k-3}{n-3-\ell\choose k-3-\ell}z^\ell\right]\mbox{d}z\ee
Since $n\geq k\geq 3,$ the sum on the right side of (\ref{hr2}) is positive for all $z\in (0,1).$ So by (\ref{om2}), $\Omega_k$ is also positive. Then by (\ref{bnu}) it follows that  $\beta_k$ is an increasing function, which proves that the strategy profile with common strategy $\beta_k$ is the unique symmetric,
increasing equilibrium of the $k$th price auction.

\vspace{0.2cm} (ii) First note that for $k=3,$ the inequalities of (\ref{bnu1}) are immediate from (\ref{k3}) (in fact, for $k=3,$ the lower bound of (\ref{bnu1}) holds with equality). So let $n\geq 4$ and
$k=4,\ldots,n.$ Taking $s=k-3,$ $r=n-3-m$ in (\ref{jen}) (Jensen's identity) for any $m>0$ we have
\be\label{hr3}\sum_{\ell=0}^{k-3}{n-3-\ell\choose k-3-\ell}z^\ell=\sum_{\ell=0}^{k-3}{m+z\ell\choose \ell}{n-3-m-z\ell\choose k-3-\ell}\ee
For $z\in [0,1],$ let $\delta(z):=(k-3)(1-z)-1$ and $\eta(z):=n-2-(k-3)z.$ If $n+4>2k,$ then $\eta(1)=n-k+1>\delta(0)+1=k-3>0.$ Fix $m\in (\delta(0)+1,\eta(1)).$
Since $\delta(z)\leq \delta(0)$ and $\eta(1)\leq \eta(z),$ for such $m,$ we have $\delta(z)<m<\eta(z)$ for all $z\in [0,1].$ We can express each term of the
right side of (\ref{hr3}) using the the gamma function as follows (note that $\Gamma(t)>0$ for all $t>0$):
$${m+z\ell\choose \ell}=\frac{\Gamma(m+z\ell+1)}{\Gamma(\ell+1)\Gamma(m-(1-z)\ell+1)}\mbox{ and }$$
\be\label{gamma}{n-3-m-z\ell\choose k-3-\ell}=\frac{\Gamma(n-2-m-z\ell)}{\Gamma(k-2-\ell)\Gamma(n-k-m+(1-z)\ell+1)}\ee
For all $\ell=0,1,\ldots,k-3,$ we have $m-(1-z)\ell+1\geq m-(1-z)(k-3)+1=m-\delta(z)>0$ and hence $\Gamma(m-(1-z)\ell+1)>0.$ We also have
$n-2-m-z\ell\geq n-2-m-z(k-3)=\eta(z)-m>0$ and hence $\Gamma(n-2-m-z\ell)>0.$ Finally $n-k-m+(1-z)\ell+1\geq n-k+1-m=\eta(1)-m>0$ and hence $\Gamma(n-k-m+(1-z)\ell+1)>0.$ This shows that
for fixed $m\in (\delta(0)+1,\eta(1)),$ {\it every term} in the sum of the right side of (\ref{hr3}) is positive for all $z\in [0,1].$ So we have
$$\sum_{\ell=0}^{k-3}{m+z\ell\choose \ell}{n-3-m-z\ell\choose k-3-\ell}=\sum_{\ell=0}^{k-3}\frac{m+z\ell}{m+z\ell}{m+z\ell\choose \ell}{n-3-m-z\ell\choose k-3-\ell}$$
$$\leq \sum_{\ell=0}^{k-3}\frac{m+z(k-3)}{m+z\ell}{m+z\ell\choose \ell}{n-3-m-z\ell\choose k-3-\ell}$$
$$=\left[\frac{m+z(k-3)}{m}\right]\sum_{\ell=0}^{k-3}\frac{m}{m+z\ell}{m+z\ell\choose \ell}{n-3-m-z\ell\choose k-3-\ell}$$
\be\label{hr4}=\left[1+\frac{z(k-3)}{m}\right]{n-3\choose k-3}<(1+z){n-3\choose k-3}\ee
where the second last equality follows by taking $s=k-3$ and $r=n-3-m$ in (\ref{hr}) (Hagen-Rothe's identity) and the last inequality follows by noting that
$m>\delta(0)+1=k-3.$ Thus (\ref{hr4}) gives an upper bound for the sum of (\ref{hr3}). Similarly we can obtain a lower bound for the sum by noting that
$$\sum_{\ell=0}^{k-3}{m+z\ell\choose \ell}{n-3-m-z\ell\choose k-3-\ell}=\sum_{\ell=0}^{k-3}\frac{m+z\ell}{m+z\ell}{m+z\ell\choose \ell}{n-3-m-z\ell\choose k-3-\ell}$$
\be\label{hr5}\geq \sum_{\ell=0}^{k-3}\frac{m}{m+z\ell}{m+z\ell\choose \ell}{n-3-m-z\ell\choose k-3-\ell}={n-3\choose k-3}\ee
where the last equality again follows by (\ref{hr}). Using the bounds from (\ref{hr4})-(\ref{hr5}), by (\ref{om2}) and (\ref{hr3}) we have
$$\frac{{n-3\choose k-3}}{\pi}\int_0^1\sqrt{\frac{z}{1-z}}\mbox{d}z\leq \Omega_k\leq \frac{{n-3\choose k-3}}{\pi}\int_0^1(1+z)\sqrt{\frac{z}{1-z}}\mbox{d}z$$
Noting that $\int_0^1\sqrt{z/(1-z)}\mbox{d}z=\pi/2$ and $\int_0^1(1+z)\sqrt{z/(1-z)}\mbox{d}z=7\pi/8,$ it follows that
\be\label{a1}\frac{1}{2}{n-3\choose k-3}\leq \Omega_k\leq \frac{7}{8}{n-3\choose k-3}\ee Since ${n-3\choose k-3}/{n-2\choose k-2}=(k-2)/(n-2),$ the result of (\ref{bnu1}) follows by applying the inequalities of (\ref{a1}) in (\ref{bnu}). \qed

\section*{Appendix}

\noi {\bf Lemma A1 (Conditional density of order statistics)} {\it Suppose $X_1,\ldots,X_m$ are iid random variables on the interval $[0,\omega],$ each having an increasing distribution function $F$ that has a continuous density $f\equiv F^\prime$ and has full support. Denote by $Y_r$ the $r$th highest order statistic of $X_1,\ldots,X_m.$ Let $x>0$ and $y\leq x.$ Then for $r=1,\ldots,m,$ the density of $Y_r$ conditional on $Y_1<x$ is
\be\label{con}h_r(y|Y_1<x)=\frac{m}{G(x)}{m-1\choose r-1}[F(x)-F(y)]^{r-1}F(y)^{m-r}f(y)\ee
where $G$ is the distribution function of $Y_1.$}

\vspace{0.1cm} \noi {\bf Proof} Denote by $H_r$ the distribution function of $Y_r$ conditional on $Y_1<x,$ that is,
\be\label{cond}H_r(y|Y_1<x)=\mbox{Pr}(Y_r\leq y|Y_1<x)=\frac{\mbox{Pr}(Y_r\leq y,Y_1<x)}{\mbox{Pr}(Y_1<x)}=\frac{\mbox{Pr}(Y_r\leq y,Y_1<x)}{G(x)}\ee
The event $(Y_r\leq y)$ is the union of $r$ mutually exclusive events as follows:
$$(Y_r\leq y)=(Y_r\leq y<Y_{r-1})\cup(Y_{r-1}\leq y<Y_{r-2})\cup\ldots \cup (Y_2\leq y<Y_1)\cup(Y_1\leq y)$$
$$=[\cup_{t=1}^{r-1} (Y_{t+1}\leq y<Y_t)]\cup(Y_1\leq y)$$
Since $y\leq x,$ we have $(Y_1\leq y,Y_1<x)=(Y_1\leq y),$ so that
$$(Y_r\leq y,Y_1<x)=[\cup_{t=1}^{r-1} (Y_{t+1}\leq y<Y_t,Y_1<x)]\cup(Y_1\leq y)$$
Hence
\be\label{x3}\mbox{Pr}(Y_r\leq y,Y_1<x)=\sum_{t=1}^{r-1}\mbox{Pr}(Y_{t+1}\leq y<Y_t,Y_1<x)+\mbox{Pr}(Y_1\leq y)\ee
Noting that
$$(Y_{t+1}\leq y<Y_t,Y_1<x)=(y< Y_j<x\mbox{ for }j=1,\ldots,t;Y_j\leq y\mbox{ for }j=t+1,\ldots,m)$$
we have
\be\label{x5}\mbox{Pr}(Y_{t+1}\leq y<Y_t,Y_1<x)={m\choose t}[F(x)-F(y)]^tF(y)^{m-t}\ee
Since $\mbox{Pr}(Y_1\leq y)=F(y)^m,$ by (\ref{cond}), (\ref{x3}) and (\ref{x5}) we have
\be\label{x4}G(x)H_r(y|Y_1<x)=\mbox{Pr}(Y_r\leq y,Y_1<x)=\sum_{t=0}^{r-1}{m\choose t}[F(x)-F(y)]^tF(y)^{m-t}\ee
Note that the conditional density $h_r(y|Y_1<x)$ is the derivative of the conditional distribution function $H_r(y|Y_1<x)$ with respect to $y.$ Also note that $F^\prime(y)=f(y).$ So from (\ref{x4}) we have
$$G(x)h_r(y|Y_1<x)=\frac{\mbox{d}}{\mbox{d}y}\sum_{t=0}^{r-1}{m\choose t}[F(x)-F(y)]^tF(y)^{m-t}$$
$$=\sum_{t=0}^{r-1}{m\choose t}[F(x)-F(y)]^t(m-t)F(y)^{m-t-1}f(y)-\sum_{t=1}^{r-1}{m\choose t}t[F(x)-F(y)]^{t-1}f(y)F(y)^{m-t}$$
{\small{$$=mf(y)\sum_{t=0}^{r-1}{m-1\choose t}[F(x)-F(y)]^tF(y)^{m-1-t}-mf(y)\sum_{t=1}^{r-1}{m-1\choose t-1}[F(x)-F(y)]^{t-1}F(y)^{m-t}$$}}
{\small{$$=mf(y)\sum_{t=0}^{r-1}{m-1\choose t}[F(x)-F(y)]^tF(y)^{m-1-t}-mf(y)\sum_{j=0}^{r-2}{m-1\choose j}[F(x)-F(y)]^jF(y)^{m-1-j}$$}}
$$=m{m-1\choose r-1}[F(x)-F(y)]^{r-1}F(y)^{m-r}f(y)$$
This completes the proof. \qed

\vspace{0.2cm} \noi {\bf Proof of Proposition 1} (i) Consider a specific bidder, say bidder $n$. By the monotonicity of $\beta$ it follows that at this strategy profile when bidder $n$ has value $x,$ it wins the object if and only if $Y_1<x,$ so $\mbox{Pr}(1\mbox{ wins})=\mbox{Pr}(Y_1<x)=G(x).$ Since $G(0)=0,$ when bidder $n$ has value $0,$ it wins with probability zero. Since a bidder makes no payment if it does not win, we conclude that $m^\beta(0)=0.$

\vspace{0.1cm} (ii) For $z\in [0,\omega],$ denote by $\wt{m}^\beta(z,x)$ the expected payment of bidder $n$ when bidder $n$ has value $x$ and it bids $b=\beta(z)$ while all other bidders follow the strategy $\beta$ (note that $\wt{m}^\beta(x,x)=m^\beta(x)$). By the monotonicity of $\beta,$ in this case bidder $n$ wins if and only if $Y_1<z,$ so $\mbox{Pr}(\mbox{bidder }n\mbox{ wins})=\mbox{Pr}(Y_1<z)=G(z).$

First let $z=0.$ Since $G(0)=0,$ in this case bidder $n$ wins with probability zero. Since a bidder makes no payment if it does not win, we have $\wt{m}^\beta(0,x)=0.$ Note that $\wt{m}^\beta(0,x)$ does not depend on bidder $n$'s value $x,$ so we have $\wt{m}^\beta(0,x)=\wt{m}^\beta(0,0)=m^\beta(0)=0.$

Next consider $z>0.$ As the auction is $k$th price, in the event bidder $n$ wins, it has to pay the $(k-1)$-th highest of the remaining bids, so we have
$$\wt{m}^\beta(z,x)=\mbox{Pr}(Y_1<z)E(\beta(Y_{k-1})|Y_1<z)+\mbox{Pr}(Y_1\geq z)0=G(z)E(\beta(Y_{k-1})|Y_1<z)$$Again observe that $\wt{m}^\beta(z,x)$ does not depend on bidder $n$'s value $x,$ so we have $\wt{m}^\beta(z,x)=\wt{m}^\beta(z,z)=m^\beta(z).$

Denote by $\pi^\beta(z,x)$ the expected payoff of bidder $n$ when bidder $n$ has value $x$ and it bids $b=\beta(z)$ while all other bidders follow the strategy $\beta.$ Since $\wt{m}^\beta(z,x)=m^\beta(z),$ we have
\be\label{pi}\pi^\beta(z,x)=G(z)x-\wt{m}^\beta(z,x)=G(z)x-m^\beta(z)\ee
We prove (ii) by using (\ref{pi}).

\vspace{0.1cm} {\bf Proof of the ``if part" of (ii)} Suppose the strategy profile where all bidders have the common strategy $\beta$ is an equilibrium. Then for   any $x\in (0,\omega),$ we must have $\pi^\beta(x,x)\geq \pi^\beta(z,x)$ for all $z\in [0,\omega].$ So the following first-order condition must hold:
$$\frac{\partial \pi^\beta(z,x)}{\partial z}[z=x]=0$$
Note from (\ref{pi}) that ${\partial \pi^\beta(z,x)}/{\partial z}=g(z)x-{\mbox{d} m^\beta(z)}/{\mbox{d} z}.$ Then by the first order condition, for all $y\in (0,\omega)$ we have $g(y)y={\mbox{d} m^\beta(y)}/{\mbox{d} y}.$ Hence
$\int_0^xyg(y)\mbox{d}y=m^\beta(x)-m^\beta(0).$ Then (\ref{m}) follows by noting that $m^\beta(0)=0.$

\vspace{0.1cm} {\bf Proof of the ``only if part" of (ii)} To prove the ``only if part", suppose the strategy profile where all bidders have the common strategy $\beta$ satisfies (\ref{m}). At this profile, for any $x\in [0,\omega],$ a bidder who has value $x$ obtains expected payoff $\pi^\beta(x,x).$ If this bidder unilaterally deviates and bids $b=\beta(z)$ for some $z\in [0,\omega],$ it would obtain $\pi^\beta(z,x).$  By (\ref{pi}) and (\ref{m}) we have
$$\pi^\beta(x,x)-\pi^\beta(z,x)=x[G(x)-G(z)]-\int_0^xyg(y)\mbox{d}y+\int_0^zyg(y)\mbox{d}y$$
Integration by parts gives $\int_0^tyg(y)\mbox{d}y=tG(t)-\int_0^tG(y)\mbox{d}y.$ So we have
\be\label{1}\pi^\beta(x,x)-\pi^\beta(z,x)=-xG(z)+\int_0^xG(y)\mbox{d}y+zG(z)-\int_0^zG(y)\mbox{d}y\ee
Note that $G$ is non-decreasing. If $z\geq x,$ then by (\ref{1}) we have
$$\pi^\beta(x,x)-\pi^\beta(z,x)=(z-x)G(z)-\int_x^zG(y)\mbox{d}y\geq (z-x)G(z)-\int_x^z G(z)\mbox{d}y=0$$
If $z<x,$ then by (\ref{1}) we have
$$\pi^\beta(x,x)-\pi^\beta(z,x)=\int_z^xG(y)\mbox{d}y-(x-z)G(z)\geq \int_z^xG(z)\mbox{d}y-(x-z)G(z)=0$$
Thus $\pi^\beta(x,x)\geq \pi^\beta(z,x)$ for all $z\in [0,\omega].$ This shows any unilateral deviation to a bid $b$ where $b=\beta(z)$ for some $z\in [0,\omega]$ is not gainful.

If a bidder with value $x$ unilaterally deviates to a bid $b<\beta(0),$ then the probability that it will win is zero and its expected payoff is also zero.
Since $\pi^\beta(x,x)\geq \pi^\beta(0,x)=0,$ such a deviation is not gainful. Finally if a bidder with value $x$ unilaterally deviates to a bid $b>\beta(\omega),$
then its expected payment is $\mbox{Pr}(\beta(Y_1)<b)E(\beta(Y_{k-1})|\beta(Y_1)<b).$ Since $b>\beta(\omega)$ and $\mbox{Pr}(Y_1<\omega)=G(\omega)=1,$ we have $\mbox{Pr}(\beta(Y_1)<b)=G(\omega)=1$ and
$E(\beta(Y_{k-1})|\beta(Y_1)<b)=E(\beta(Y_{k-1})|Y_1<\omega).$ This shows its expected payment is $m^\beta(\omega)$ and expected payoff is $\pi^\beta(\omega,x).$
Since $\pi^\beta(x,x)\geq \pi^\beta(\omega,x),$ such a deviation is also not gainful.

This proves that if an increasing, symmetric strategy profile with common strategy $\beta$ satisfies (\ref{m}), then this profile is an equilibrium. \qed

\vspace{0.2cm} The following lemma will be useful to prove Lemma 1.

\vspace{0.2cm} \noi {\bf Lemma A2} {\it Let $n\geq k\geq 3.$ For $t=2,\ldots,k$ and $\ell=0,1,\ldots,k-t,$ let $\lambda_{\ell,t}:={k-t\choose \ell},$}
\be\label{g1}\gamma_{\ell}(x):=\int_{0}^{x} \beta_{k}(y)F(y)^{n-k+\ell}f(y)\mbox{d}y,\Phi_t(x):=\sum_{\ell=0}^{k-t}(-1)^\ell \lambda_{\ell,t}F(x)^{k-t-\ell}\gamma_{\ell}(x)\ee
{\it Then for $t=2,\ldots,k-1,$ the following hold}: $\Phi^\prime_t(x)/f(x)=(k-t)\Phi_{t+1}(x).$

\vspace{0.2cm} \noi {\bf Proof} Note that $F^\prime(x)=f(x).$ Also note that since $\gamma^\prime_{\ell}(x)=$ $\beta_{k}(x)F(x)^{n-k+\ell}f(x),$ we have $F(x)^{k-t-\ell}\gamma^\prime_{\ell}(x)/f(x)$ $=\beta_{k}(x)F(x)^{n-t}.$
Using these, by (\ref{g1}) we have
$$\frac{\Phi^\prime_t(x)}{f(x)}=\sum_{\ell=0}^{k-t-1}(-1)^\ell \lambda_{\ell,t}(k-t-\ell)F(x)^{k-t-\ell-1}\gamma_{\ell}(x)$$$$+
\beta_k(x)F(x)^{n-t}\sum_{\ell=0}^{k-t-1}(-1)^\ell \lambda_{\ell,t}+(-1)^{k-t}\lambda_{k-t,t}\beta_k(x)F(x)^{n-t}$$
Since $(k-t-\ell)\lambda_{\ell,t}=(k-t)\lambda_{\ell,t+1}$ the expression above equals
$$(k-t)\sum_{\ell=0}^{k-t-1}(-1)^{\ell}\lambda_{\ell,t+1}F(x)^{k-t-1-\ell}\gamma_{\ell}(x)+\beta_k(x)F(x)^{n-t}\sum_{\ell=0}^{k-t}(-1)^\ell\lambda_{\ell,t}$$
The result follows by noting that the first sum above is $(k-t)\Phi_{t+1}(x)$ and the second sum is zero. \qed

\vspace{0.2cm} \noi {\bf Proof of Lemma 1} Recall that
$$\phi_0(x):=\int_{0}^{x} \beta_k(y)[F(x)-F(y)]^{k-2}F(y)^{n-k}f(y)\mbox{d}y$$So we have
$$\phi_0(x)=\int_{0}^{x} \beta_{k}(y)\left[\sum_{\ell=0}^{k-2}(-1)^\ell {k-2\choose \ell}F(y)^\ell F(x)^{k-2-\ell}\right]F(y)^{n-k}f(y)\mbox{d}y$$
Changing the order of summation and integration, noting that $\lambda_{\ell,t}={k-t\choose \ell},$ and using the functions $\gamma_{\ell}(x),\Phi_t(x)$ from (\ref{g1}), we have
$$\phi_0(x)=\sum_{\ell=0}^{k-2}(-1)^\ell \lambda_{\ell,2}F(x)^{k-2-\ell}\gamma_{\ell}(x)=\Phi_2(x)$$
Using Lemma A2, we have
$\phi_1(x)=\phi^\prime_0(x)/f(x)=\Phi^\prime_2(x)/f(x)=(k-2)\Phi_3(x).$ Again applying Lemma A2: $\phi_2(x)=\phi^\prime_1(x)/f(x)=(k-2)\Phi^\prime_3(x)/f(x)=(k-2)(k-3)\Phi_4(x).$ Using this reasoning after
$k-2$ steps gives
$$\phi_{k-2}(x)=(k-2)\times \ldots \times 1\times \Phi_k(x)=(k-2)!\Phi_k(x)$$Note by (\ref{g1}) that $\Phi_k(x)=\gamma_{0}(x)=\int_{0}^{x} \beta_{k}(y)F(y)^{n-k}f(y)\mbox{d}y.$
This implies $\Phi^\prime_k(x)=\gamma^\prime_0(x)=\beta_{k}(x)F(x)^{n-k}f(x).$ Hence $\phi_{k-1}(x)=\phi^\prime_{k-2}(x)/f(x)=(k-2)!\Phi^\prime_k(x)/f(x)=(k-2)!\beta_{k}(x)F(x)^{n-k}.$ \qed

\section*{Acknowledgements}

For helpful comments and suggestions, we are most grateful to an anonymous reviewer and Tsogbadral Galaabaatar.

\end{document}